# Graph State Generation Based on Quantum Photonic Devices, Enabling Measurement-Based Quantum Computing

Masoud Hakimi Heris

*Abstract*—**Photonic cluster states are fundamental resource for measurement-based quantum computing (MBQC), which is a promising approach for scalable quantum computing. These highly entangled states enable computation with applying local measurements rather than gate operations, which makes them essential for fault-tolerant quantum computing. Generation of the photonic cluster states requires high-quality entanglement and stability, which can be done by making use of cavity quantum electrodynamics (cavity QED), which is embedding quantum dots (QDs) into photonic microcavities. This method enables controlled single-photon emission and the entanglement of them, helping with key challenges in generating photonic cluster states.**

**Enhancing the generation of the photonic cluster state requires improvement in the quality factor (Q-factor) and therefore, cooperativity of cavity and QD systems. A high Q-factor helps with reduction in photon loss and decoherence, ensuring more stable and high-fidelity quantum states. It also enables better light-matter interactions, improving cluster state quality. Cooperativity, which is a measure for the light-matter coupling strength relative to dissipation, directly impacts photon emission efficiency. Optimization of these parameters improves entanglement generation, which is essential for producing high-quality photonic cluster states. Therefore, optimized cavities are essential for improving the generation of the photonic cluster states. Scalability is also an important parameter for the generation of the large-scale cluster states used in quantum computing. In this paper, after reviewing MBQC, cluster states, and the generation of the photonic cluster states, we investigate the deterministic generation of the photonic cluster state resource for the purpose of MBQC, and highlights the importance of the inverse design techniques for this goal.**

## I. INTRODUCTION

IN 1935, Einstein, Podolsky, and Rosen (EPR) brought up a paradox which questioned the completeness of quantum mechanics, arguing that "spooky action at a distance" conflicted with local realism [1]. Schrödinger, in response, introduced the term entanglement and identified it as a fundamental feature of quantum theory [2]. In 1964, John Bell formulated his famous theorem, indicating that no local hidden variable theory could completely explain quantum correlations [3]. Decades later, experiments by Alain Aspect [4], John Clauser [5], and Anton Zeilinger [6], each contributed for violations of Bell’s inequalities, meaning a strong confirmation of the reality of entanglement. Their groundbreaking efforts resulted in the 2022 Nobel Prize in Physics, further proof of their contributions to the field of quantum mechanics. This achievement established entanglement as a core principle of quantum mechanics, challenging classical beliefs about how the nature works and paving the way for modern quantum technologies.

Entanglement can be considered a the heart of quantum cryptography, making secure communication possible in a way that classical systems never could make it possible. It plays an essential role in protocols like BB84 [7], where the no-cloning theorem and collapse of the wavefunction after measurement can be exploited to make an unbreakable encryption. Entanglement is also a key feature in quantum algorithms such as Shor’s algorithm [8], which is for factoring large numbers exponentially faster than classical methods. It is crucial for performing the calculations that would take infinitely large amount of time for traditional computers. Moreover, entanglement enables other exciting applications like quantum teleportation, where the information is distributed instantaneously across long distances, and quantum networks, which could one day form the foundation of a completely new form of the internet.

In 2001, Briegel and Raussendorf introduced a new way of quantum computing called Measurement-Based Quantum Computing (MBQC) [9]. On the contrary to the gate-based quantum computing, which is based on a sequence of logic gates being applied to qubits, MBQC works on a highly entangled state, known as a cluster state, and then computes the entire algorithm through a series of measurements. Measurement patterns govern the algorithm. This was fundamentally different from the previous way of processing quantum information. It is also promising for photonic quantum computing, for which generation and measuring entangled photons seem easier than maintaining stable quantum gates [10]. The higher dimensions of cluster states also makes MBQC a strong promise for fault-tolerant quantum computing, as it can help with resisting over errors and decoherence. MBQC could be considered as the practical promise in making large-scale and fault-tolerant quantum computing [10].

Research in this field has been focusing on generating and manipulating complex pattern entangled states, such as different graph states, where qubits are represented as vertices and entanglement as edges [11]. In MBQC, a two-dimensional cluster state is created in the first place, and computation takes place through sequential single-qubit measurements in specific bases and in different patterns depending on the algorithm to be implemented. The choice of qubit and base of measurement and sequence of these measurements determine the entire algorithm [12]. Among all qubit platforms, optical photons areexcellent qubit carriers since they experience very little decoherence and work at room temperature [13].

The most common method for generating such entangled photons is called spontaneous parametric down-conversion (SPDC), which is a probabilistic approach. The probabilistic nature of this approach makes it difficult to create larger states [10]. To tackle this limitation, deterministic schemes have been proposed [14], where a single-spin memory qubit handles the entanglement on a sequence of emitted photons. This method is efficient, as it allows the generation of infinite number of entangled photons from a single device [15].

In this paper, following a brief review of photonic graph state generation and their applications, we investigate strategies to enhance the efficiency and fidelity of spin-photon quantum systems in generating large-scale cluster states. Specifically, we focus on improving the cooperativity of cavity quantum electrodynamics systems by optimizing the cavity quality factor and far-field emission profile through advanced nanophotonic design techniques.

## II. Graph States and MBQC

The definition of a graph states which depicts different patterns of entanglement between multiple qubits can be represented by the mathematical concept of a graph. Figure 1 [12] illustrates three types of graphs. A graph consists of a set of vertices and edges that connects them. In order to create a quantum graph state, each vertex is replaced with a qubit, and entanglement is applied between any two qubits connected by an edge. One of the important types of highly entangled graph state is called cluster state (represented in Figure 1 (a) and (b)), where the graph structure looks like a 1D, 2D, or even 3D lattice structure. Cluster states are the fundamental resource in measurement-based quantum computing (MBQC). In this field, entangled states, or actually the cluster states, initially gets prepared as the resource, and then the quantum gates or processing operations are implemented through different patterns of local measurements [11].

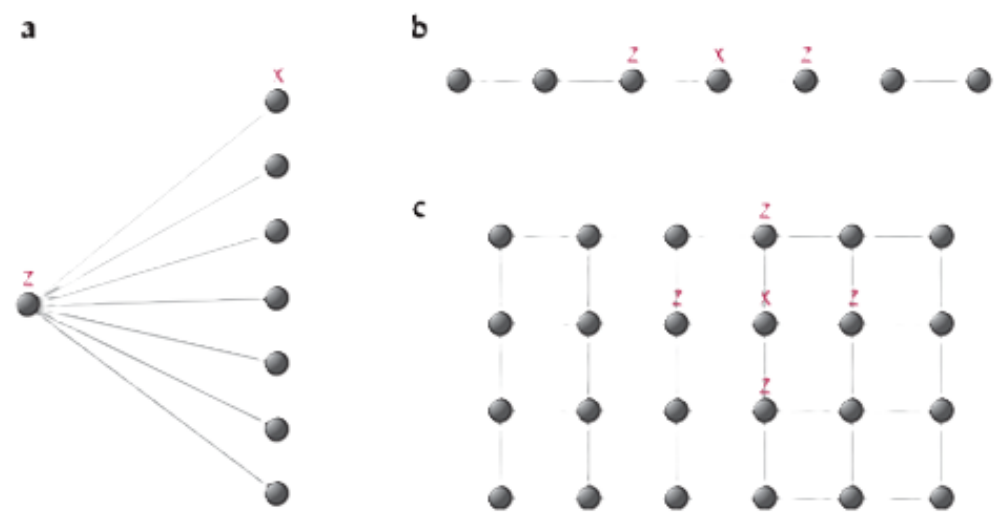

**Fig. 1.** Different types of graphs states, (a) GHZ graph states, (b) 1D and (c) 2D cluster states [12].

The first demonstration of measurement-based entanglement of macroscopically individual systems was reported by Chou et al. in 2005 [16], using atomic ensembles. Later, in 2007, Moehring et al. reported the entanglement of two remote single atoms [17].

Figure 2 illustrates the MBQC process, showing how different measurement patterns on a 2D cluster state of entangled qubits can implement a quantum operation, such as realizing quantum gates on other qubits. MBQC is universal, although the measurement outcomes at each step of the computation are random, due to the no information about the state of the qubits, any quantum computation can be deterministically realized [18]. It is noteworthy that in MBQC we need more qubits compared to that of the gate-based quantum computing, the reason being that we need additional qubits to help us with the computation process other than the qubits holding the information.

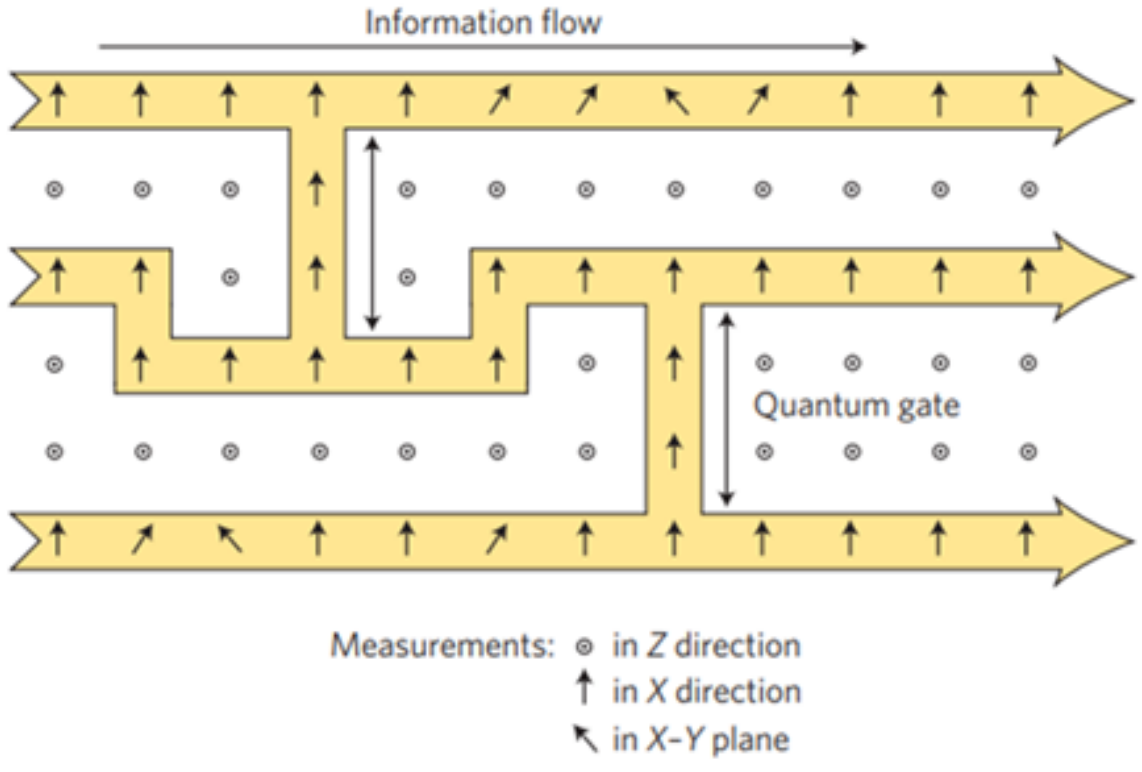

Fig. 2. Quantum computation by measuring qubits in differene bases on a lattice. Before the measurements the qubits are in the 2D cluster state. Circles symbolize measurements of Z direction, vertical arrows are measurements of X direction, while tilted arrows refer to measurements in X-Y plane [12].

In the first step of the computation, a subset of qubits is measured in the basis of Z which effectively removes them. In Figure 2 these qubits are denoted by circles with a dot in the middle. The yellow pathways in Figure 2 illustrate the flow of information, where arrows represent different measurement bases (X, and Y Pauli matrix eigenvalue bases). The section marked "Quantum gate" depicts a region where a logical operation is performed through measurement, a CNOT gate here for instance, showcasing how MBQC performs unitary transformations without using any physical gate implementations [12]. The measurements in Y basis also are representing the single qubit rotation. All these implementations through the measurements in different bases are actually the required ingredients of quantum computation.

Figure 3 shows a practical example of MBQC, specifically showcasing the realization of a CNOT gate through single-particle measurements [12]. In Figure 3 (a), a simplified setup for implementing the CNOT gate is shown. The "target in" and "target out" qubits represent the target qubit’s initial and final states, respectively, while the "control" qubit influences the target through the entangled state. The arrows indicate measurements in various bases, and the results of these measurements determine the outcome of the computation.

Figure 3 (b) further demonstrates this by showing a cluster-state-based implementation of the CNOT gate in MBQC. The horizontal and vertical connections demonstrate how the entanglement structure enables control-target interactions which is essentially CNOT gate. The measurement pattern ensures that quantum correlations between the control and target qubits produce the desired logical process, which depends on the measurement outcomes. Adaptive measurements, that subsequent measurements depend on the previous results, are actually for correctly implementing the CNOT operation [9].

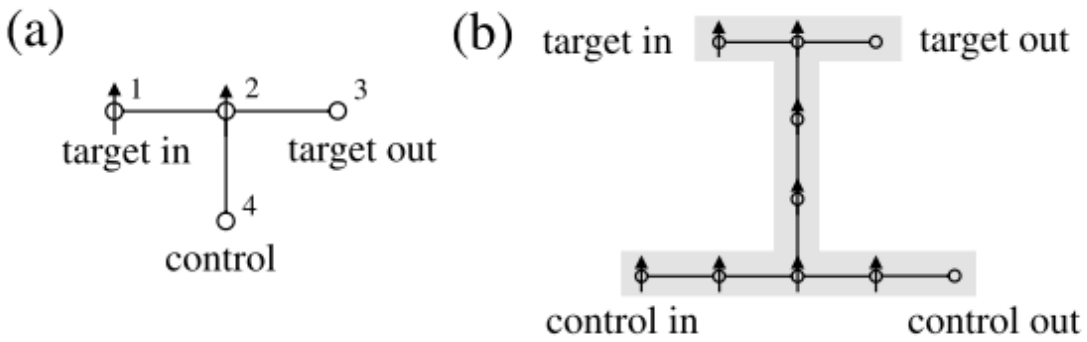

**Fig. 3.** Realization of CNOT gate using MBQC [9].

Therefore, in order to conclude, applying quantum operations takes place through a systematic procedure: First, a cluster state is generated. Then, individual qubits are measured in chosen pattern and bases, with the measurement results leading to next measurements. Finally, classical post-processing operations should be implemented to control and modify the outcomes based on previous measurements, effectively performing logical operations on the remaining unmeasured qubits [18].

As can be understood, entanglement plays a crucial role in MBQC. This approach is a universal procedure of computation because every quantum operation can be implemented, in addition it reduces the need for intricate gate-based interactions, instead depending on pre-established quantum correlations, i.e. cluster states [9].

Another important capability of MBQC is its built-in adaptability for fault-tolerant quantum computing through the planning subsequent measurements. Because of the dependence of computations on single-qubit measurements, errors can be addressed using an extra dimension in the cluster state called error-correcting cluster states, where redundant encoding and measurements improve reliability [12]. This makes MBQC a strong candidate for scalable quantum computing, particularly photonic quantum computing, where creating large-scale of photonic entanglement is achievable.

In terms of implementation of photonic MBQC, entangled photonic states can be produced through parametric down-conversion or other nonlinear optical processes, and single-photon measurements which offer a natural method for performing the required quantum operations. Recent experimental demonstrations have successfully proved small-scale MBQC systems using photonic cluster states, promising its practical possibility [10].

Photonic MBQC offers a unique and powerful approach to quantum computation, where entangled resource states and measurement sequences replace conventional unitary gate operations. According to the growing number of research, photonic MBQC is anticipated to be great opportunity in the development of large-scale fault-tolerant quantum computers.

## III. Deterministic Genration of Photonic Cluster State

Entanglement in photonic cluster states can be implemented exploiting various degrees of freedom of photons, such as polarization, time-frequency, and spatial modes of photons [19]. Single and entangled photons can be generated by the spontaneous emission of an optical transition from an atom or an atom-like quantum photon emitter [20]. Artificial atoms, such as semiconductor quantum dots (QDs), have been exceptional quantum emitters due to their ability of integration into electro-optical devices [20]. The optically active quantum dots can emit photons at extremely high rates, in the range of less than a nanosecond, and high indistinguishability. Both of these features are necessary for the generation of stable and reliable cluster state resources [21].

The first realization of photonic cluster states that were conducted using quantum dots, demonstrated entanglement of up to three and four photonic qubits in a linear cluster state [15]. Main challenges in creating higher photon numbers suffered from low photon generation rate, low collection efficiencies, noisy semiconductor environment, and the need for a probabilistic entangling gate which needed to be addressed in the research [15].

There are two approaches to generate entangled photon states using single-photon emitters. In the first approach, single photons are generated separately and then entangled using a beamsplitter [22]. The main challenge of this method is that the entanglement generation between photons is probabilistic, which is a big drawback. Therefore, due to this probabilistic nature, the successful number of entangled photons decreases greatly, as the number of photons increases, which limits the scalability and hence the potential for photonic quantum computing [15].

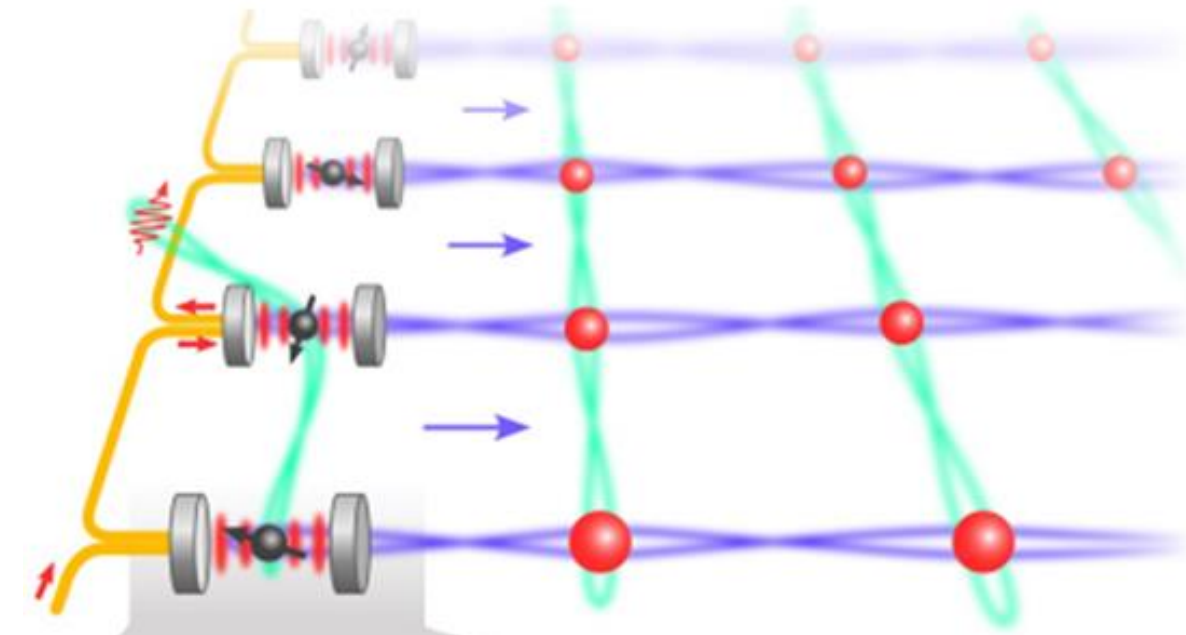

**Fig. 4.** Each source produces a one-dimensional photonic cluster state while a two-dimensional cluster states is woven from the individual 1D cluster states [15].

In the second approach however, which was proposed by Lindner and Rudolph [23], generating one-dimensional photonic cluster states using only semiconductor quantum dots (QDs) are deterministic. Their approach utilizes a single QD's confined electron or hole spin precessing under an external magnetic field while it is driven by a sequence of laser pulses with different durations. After the QD gets excited, due to the relaxation of excited electron, a single photon is deterministically emitted. The polarization of the emitted photon depends on which state it relaxes back to, resulting in the photon polarization entangled with the state of the QD-confined spin [20]. This process is then repeated multiple times to generate a large 1D cluster state of entangled photons. Further, in order to obtain a 2D cluster state, two strings of 1D cluster states from different single-photon sources can be woven or fused using fusion techniques, as shown in Figure 4 [15]. Although the fusion process is also probabilistic, there are other complicated approaches proposed to generate 2D cluster states from a single-photon source through more complex gates and processes [24].

To better understand how Lindner and Rudolph's [25] idea works, let's dive into an example of realization through the insightful work by Dan Cogan et al. [13]. Figure 5 illustrates the experimental realization of the circuit for generating one-dimensional cluster states proposed by them. In this example, an on-demand, continuously generated string of indistinguishable photons entangled in a cluster state is represented. They have been inspired by the idea of Gerardot, Brian D., et al. [26], to optically manipulate the hole spin in a QD. The heavy-hole (HH) spin is leveraged in this realization as the entangler. The fact that after photon emission, HH is left in a stable ground state makes the emitted single photons highly indistinguishable. The HH has a $^{1}/_{2}$ spin, and in the absence of an external magnetic field, its two spin states are degenerate. This causes the precession rate to vanish if there is no external magnetic field. Hence, by finely tuning of the external field's strength, it lifts the degeneracy and allows control over the spin precession and consequently the photon emission rates which improves entanglement robustness in the cluster state [13].

Figure 5 (a) illustrates a schematic of periodic sequence of controlled-not (CNOT) and Hadamard gates (using laser pulses) is applied to the HH of the QD to produce the cluster state. Furthermore, In Figure 5 (b) the schematic of the QD-based device and the experimental system for generating and characterizing the cluster state is shown. Each laser pulse (red upward arrows) excites the confined HH (|h⟩ = |⇑⟩) to the excited trion state represented by Trion* (|T∗⟩ = |⇑⇓↑⟩*), where an electron is excited and leaving another hole behind creating a trion. The excited electron then decays back to the trion ground state (|T⟩) in about 5 ps which causes emission a spin-preserving optical phonon. The trion state then recombines after approximately 400 ps, that emits a photon (represented by pink arrow) and thus leaving the HH residing in its ground state (|h⟩). A dichroic mirror which separates transmission and reflection based on the wavelength directs the emitted photons to the detectors. Dichroic mirror is used to separate the excitation and collection lights.

The heavy-hole (HH) state and the positive trion state help us realizing a spin qubits. Relaxation of the excited state to each of the ground states, which are separated by the external magnetic field, determined the polarization of the emitted photon. Therefore, the polarization of the emitted photon is entangled with the spin of the QD. The relaxation path picking is called selection rule, which is actually the fundamental of entanglement generation is this device. In another point of view each of the excitation-emission steps are the realization of a CNOT gate between the spin and the photon qubits. Moreover, the Hadamard gate on the spin qubit is also realized by adjusting the time of the excitations and the field-induced precession of the HH using the external magnetic field, which is essentially putting the spin qubit is superposition. Therefore, by a sequence of laser pulses and making use of an external magnetic field, deterministic generation of the cluster state can be taken place. After each pulse sequence an entangled photon will be added to the cluster state [20].

Schwartz et al. (2016) [27], successfully generated a 1D cluster state of five entangled photons long using a confined Dark Exciton (DE) technique. Vezvaee et al. (2022) [28] employed QD molecules in order to deterministically generate photonic cluster states with both a high production rate and a fidelity of more than double of its previous works. Cogan et al. in their 2021 research [29] and their later work in 2023 [20] achieved longer 1D cluster states of 10 entangled photons with high indistinguishability. Regarding multiple dimension photonic cluster states however, Shi and Waks (2021) [24] proposed a protocol to deterministically generate multi-dimensional photonic cluster states using a single atom-cavity system and making use of the time-delay feedback. Furthermore, Ferreira et al. (2024) [30] generated a 2D cluster state of four photons with 70% fidelity.

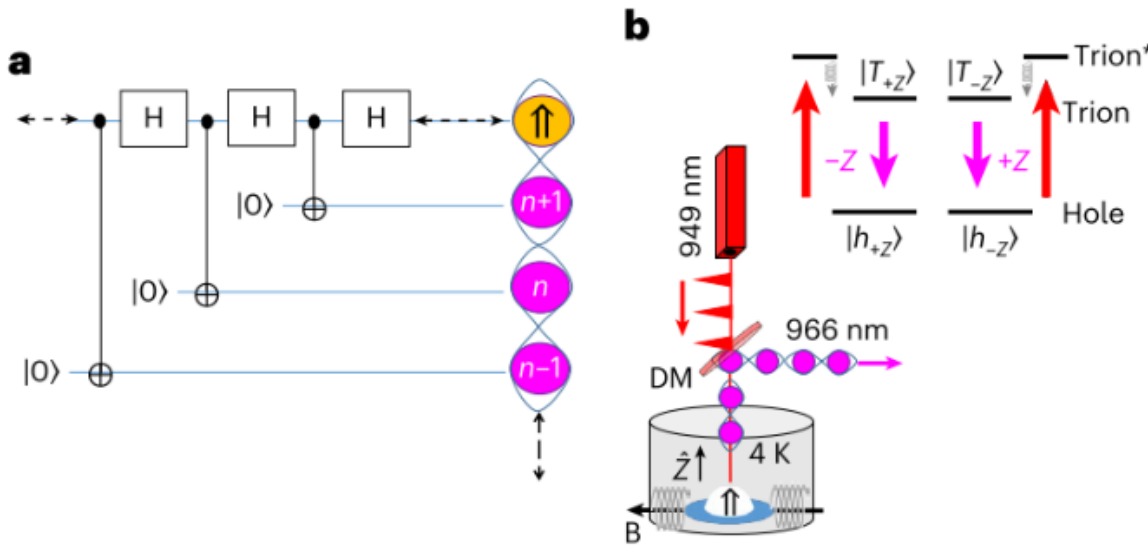

**Fig. 5.** (a) Circuit realization of the cluster state generation using sequential Hadamard and CNOT gates. (b) A schematic of the QD-based device. DM represents a dichroic mirror and} %B ⃗ = B$\hat{x}$ the externally applied magnetic field. The inset shows polarization selection rules of optical transition [20].

## IV. Gap Analysis

All the great experimental works mentioned in the end of previous section suffer from the use of high quality cavities. Integrating high-Q bullseye cavities into such emitters can significantly enhance photon generation efficiency and coherence, which will help with generation of larger and more robust cluster states, like the outstanding work of Coste, N., et al. [31]. High-Q cavities could improve light-matter interaction, wind up stronger photon entanglement and reduce decoherence effects, which are essential to scale up the size cluster states. Making use of the optimized cavity designs in order for achieving Q factor cavities could enhance single-photon emission rates and indistinguishability, which helps with generating longer 1D and 2D cluster states. This enhancement could improve the photon numbers over the current limits, making large-scale photonic quantum computing possible.

The fidelity of the achieved CNOT gates is influenced by the spin-photon interface cooperativity parameter which is shown in equation (1):

$$C = {}^{2g^2}/_{\kappa\gamma}, \tag{1}$$

where g represents the coupling strength between the spin system and the cavity, κ and γ are the photon loss from the cavity and the loss of the spin system, respectively [24].

In addition to the high Q factor, optical accessibility of the cavity also matters a lot. Although, Photonic Crystal cavities can achieve very high quality factors but they extremely suffer from low optical accessibility due to their divergent far-field emission pattern. Amongst all photonic cavity types, circular gratings_ bullseye cavities_ are promising for having good optical accessibility as they have near gaussian far-field emission pattern and support circular polarization due to their symmetrical geometry.

An efficient spin photon interface which can simultaneously achieve high cooperativity, support efficient coupling to light and circular polarization, can be realized by coupling the dots to nanofabricated bullseye cavities. According to a study by Harjot Singh et al. [32], the conventional periodic bullseye cavities have demonstrated over 60% optical access efficiency, enabled the transmission of circularly polarized light to the quantum dot. Figure 6 (a) demonstrates the intensity of laser light reflected from the periodic bullseye cavity as a function of the spectral detuning of the laser from the optical transition of the dot. By fitting the simulated spectrum (red solid curve) with the experimental results (blue dots), the parameters of the system, g=35 GHz, and $\gamma$=1 GHz, are extracted. Using equation 1, a cooperativity as high as C≈8 is achieved, which corresponds to an efficiency of over 80% in flipping the polarization of a photon. Figure 6 (b) shows the SEM image of the fabricated cavity, and Figure 6 (c) depicts the simulated far-field emission pattern of the same cavity.

Since the Q factor of the conventional periodic bullseye cavities are limited, and the cooperativity is proportional to the Q factor, to improve the cooperativity, the bullseye cavities with higher Q factors need to be designed. Therefore, exploiting optimization and inverse design techniques to design high-Q bullseye cavities can pave the way for the next-generation single photon emitters.

## V. Optimization and Design of Bullseye Cavity

Coupling optically active spin systems to photonic cavities with high cooperativity enables efficient light-matter interactions, which are fundamental to the operation of quantum networks and quantum information processing. Bullseye cavities, consisting of concentric rings etched into a substrate, are a promising design as explained in section IV. However, conventional circular gratings with constant period, satisfying the Bragg scattering condition have limited quality factors (1000), leading to low cooperativity [32].

Using inverse-design methods, non-periodic circular cavity geometry can be optimized to maximize the key figures of merit, namely, quality factor, optical access through Gaussian far-field emission pattern, and polarization degeneracy. In contrast to conventional intuition-driven design approaches, inverse design techniques use computational algorithms to efficiently optimize photonic structures for desired optical characteristics. There are two primary inverse-design approaches including machine learning (ML) and computational optimization algorithms. ML uses models that have been trained on data calculated by simulations, for the generation of optimized designs and therefore extra simulations are not required. In contrast, computational optimization algorithms are based on improving and refining the designs iteratively through several simulation steps.

Optimization algorithms can be classified into gradient based (Automatic Differentiation (AD), Adjoint Method (AM)) and non-gradient based methods (Genetic Algorithm (GA), Particle Swarm optimization (PSO)). Gradient-based optimization methods use gradient information for differentiable functions to find optimal solutions. In contrast to gradient-based approaches which struggle to find the global solutions, non-gradient methods explore the solution space with heuristic techniques or random search strategies, making them a powerful tool for more complex problems as they do not get stuck in a local optimum and have the ability of finding the global optimum.

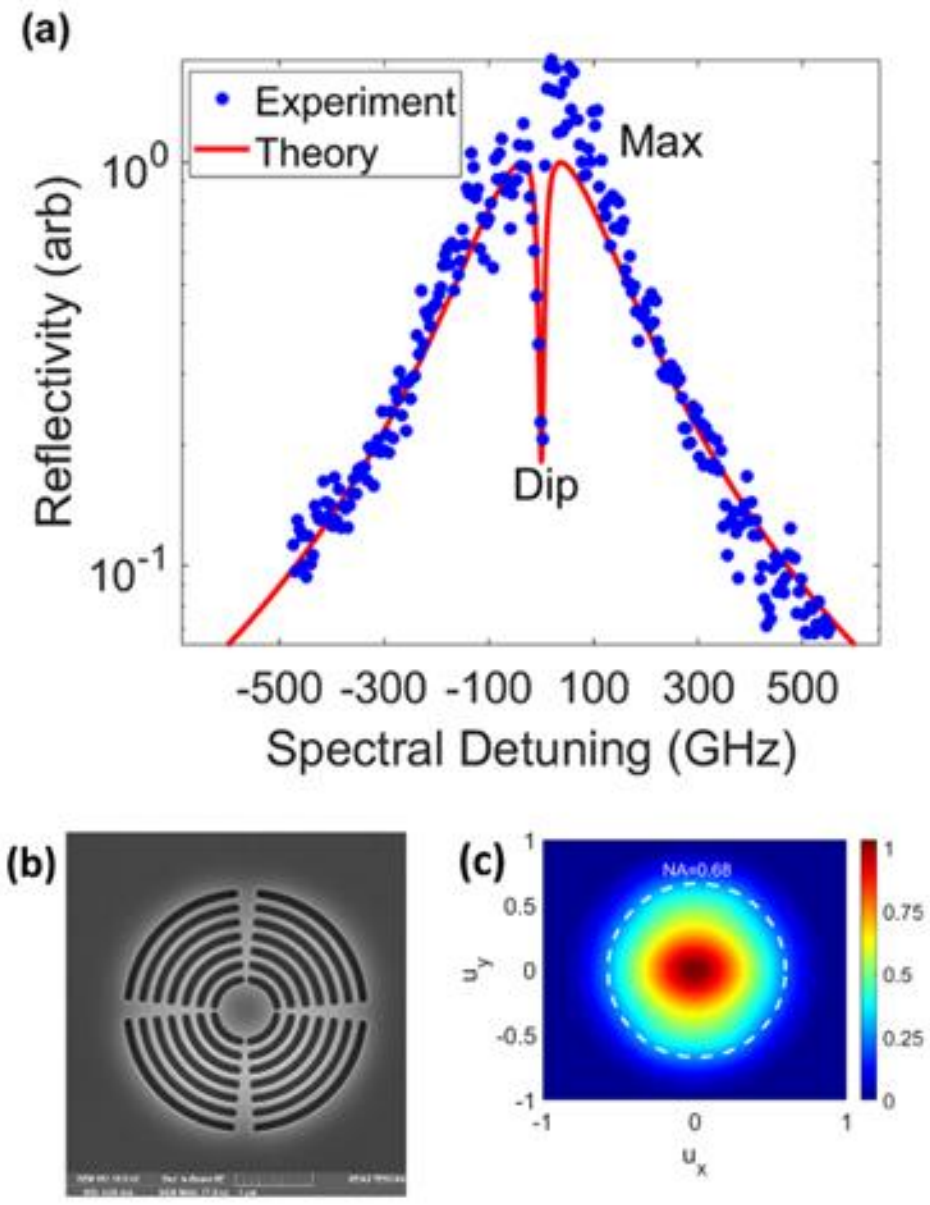

**Fig. 6.** (a) Intensity of laser light from the bullseye cavity coupled the QD emitter as a function of the spectral detuning of the laser from the optical transition of the dot. (b) SEM image of fabricated cavity and (c) FDTD simulation of the far-field emission pattern of the designed bullseye cavity from [32].

GA, a non-gradient-based optimization technique, inspired by the principles of natural selection and evolution, is particularly well-suited for addressing complex, high-dimensional optimization challenges where traditional gradient-based methods may fall short. One notable application of GA in photonics is the optimization of the quality factor (Q-factor) in optical cavities [33].

While the traditional bullseye cavity design methods often rely on intuition or parameter sweeps, which become computationally expensive as design complexity increases, GA overcomes this challenge by efficiently exploring the parameter space, refining structures iteratively based on fitness functions that assess metrics such as Q-factor, mode volume, or other performance indicators [33]. Research has shown that GA-optimized cavity geometries can significantly enhance Q-factors beyond traditional designs by fine-tuning the perturbation of the photonic bandgap [34].

In our latest efforts, we have proposed and numerically validated a non-periodic bullseye cavity optimized using Genetic Algorithms (GA), building on the previous work in ref [35] on a periodic bullseye cavity. The optimized cavity achieves a quality factor exceeding 5000, five times greater than that of the conventional periodic cavity, and operates at 944 nm, which is within the emission range of InAs/GaAs quantum dots (910 nm- 970 nm). Figure 7 (a) shows an SEM image of our currently fabricated periodic bullseye cavity based on dimensions of ref [32] which is a base for this project to pursue on. The dark areas in the SEM image represent the etched areas on the GaAs substrate and the vertical and horizontal tapered lines represent "bridges" that enable the electrical charging of the dots and prevent the suspended structure from collapsing. Figure 7 (b) shows the spectrum of the optimized non-periodic cavity with resonance wavelength of 944 nm. Additionally, the cavity exhibits a near-Gaussian far-field emission pattern, providing easy and efficient optical access. Figure 7 (c) shows the far-field emission pattern of the optimized cavity found by finite-difference-time-domain (FDTD) simulations using Lumerical. Comparing the results of our design with that of the [32] as shown in Figure 6 (c) the polarization independent nearly Gaussian emission pattern can be collected using an objective lens with numerical aperture of NA=0.68, thereby highlighting the potential of the cavity in efficiently exciting and collecting photons from quantum dots coupled to it. The polarization degeneracy positions the bullseye cavities as being promising for the coherent control of the spin of electrically charged quantum dots coupled to the cavity, which typically requires circularly polarized light. Moreover, it is noteworthy to mention that we have designed an efficient optical setup to efficiently excite and collect the light from the cavity, as illustrated in Figure 8 presents a 2D schematic of the optical setup, while the inset illustrates its 3D configuration.

## VI. Conclusion

The optimization of the bullseye cavity, achieving a quality factor exceeding 5000, represents a significant step forward in the efficient generation of photonic cluster states. A high Q-factor reduces photon loss, suppresses decoherence, and enhances cooperativity, thereby improving the interaction between single-photon emitters and cavities. These improvements are crucial for generating large-scale, high-fidelity cluster states, which form the backbone of measurement-based quantum computing (MBQC). By strengthening emitter-cavity coupling, we ensure efficient photon extraction and entanglement, key to scalable quantum processing. Additionally, higher-Q cavities contribute to improved coherence times, allowing quantum information to be preserved for longer durations, which is essential for fault-tolerant quantum operations. As a result, these enhancements not only benefit MBQC but also strengthen the broader field of quantum optics by enabling more efficient sources of indistinguishable photons, crucial for applications such as quantum key distribution and linear optical quantum computing.

Beyond the immediate benefits to photonic cluster state generation, these advancements pave the way for more robust and practical implementations of MBQC. Higher-Q cavities enable stronger light-matter interactions, facilitating deterministic and coherent photon emission, which is essential for integrating quantum networks and logic gates. The ability to achieve such strong interactions with minimal loss is fundamental for implementing high-fidelity quantum gates and long-distance quantum communication. Moreover, the improvement in cavity designs can lead to the development of more versatile photonic architectures, including hybrid quantum systems that integrate solid-state emitters, superconducting circuits, or atomic systems. Further refinements in cavity design, material engineering, and coupling mechanisms could push these improvements even further, bringing us closer to fault-tolerant and scalable photonic quantum computing. As we continue to optimize emitter-cavity systems, the prospect of realizing large-scale quantum networks and processors becomes increasingly feasible, unlocking new possibilities in quantum information science. These advancements not only strengthen fundamental quantum research but also accelerate progress toward practical quantum technologies, making quantum computing and secure communication more accessible in the near future.

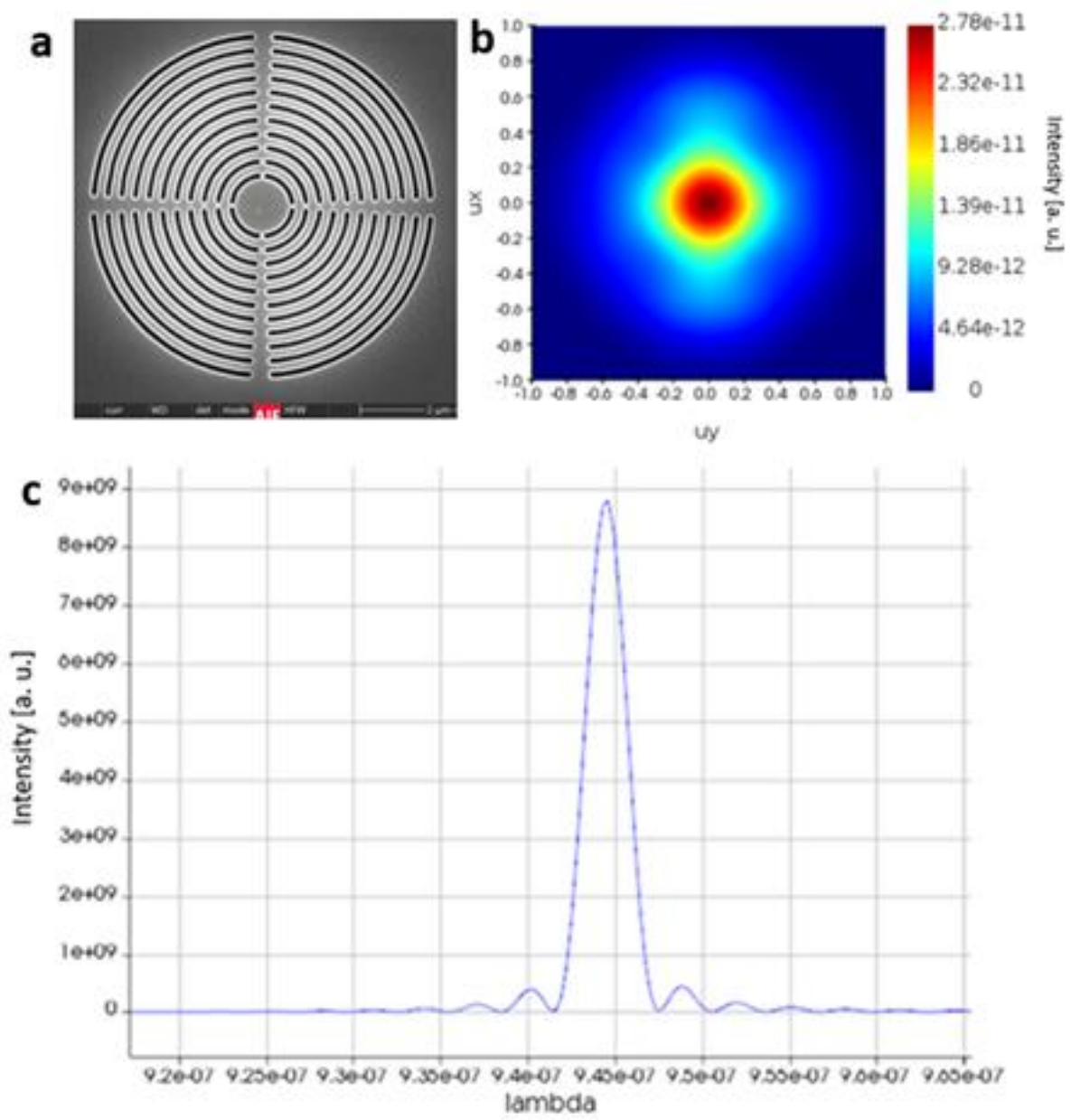

**Fig. 7.** (a) Fabricated cavity based on the reference [32] at NC State University. (b) Near gaussian far-field emission of the optimized cavity. (c) Designed optical setup for excitation and detection of light to and from the single photon emitter. (d) Emission intensity spectrum of the optimized cavity representing the enhancement at 944 nm with over 5000 quality-factor.

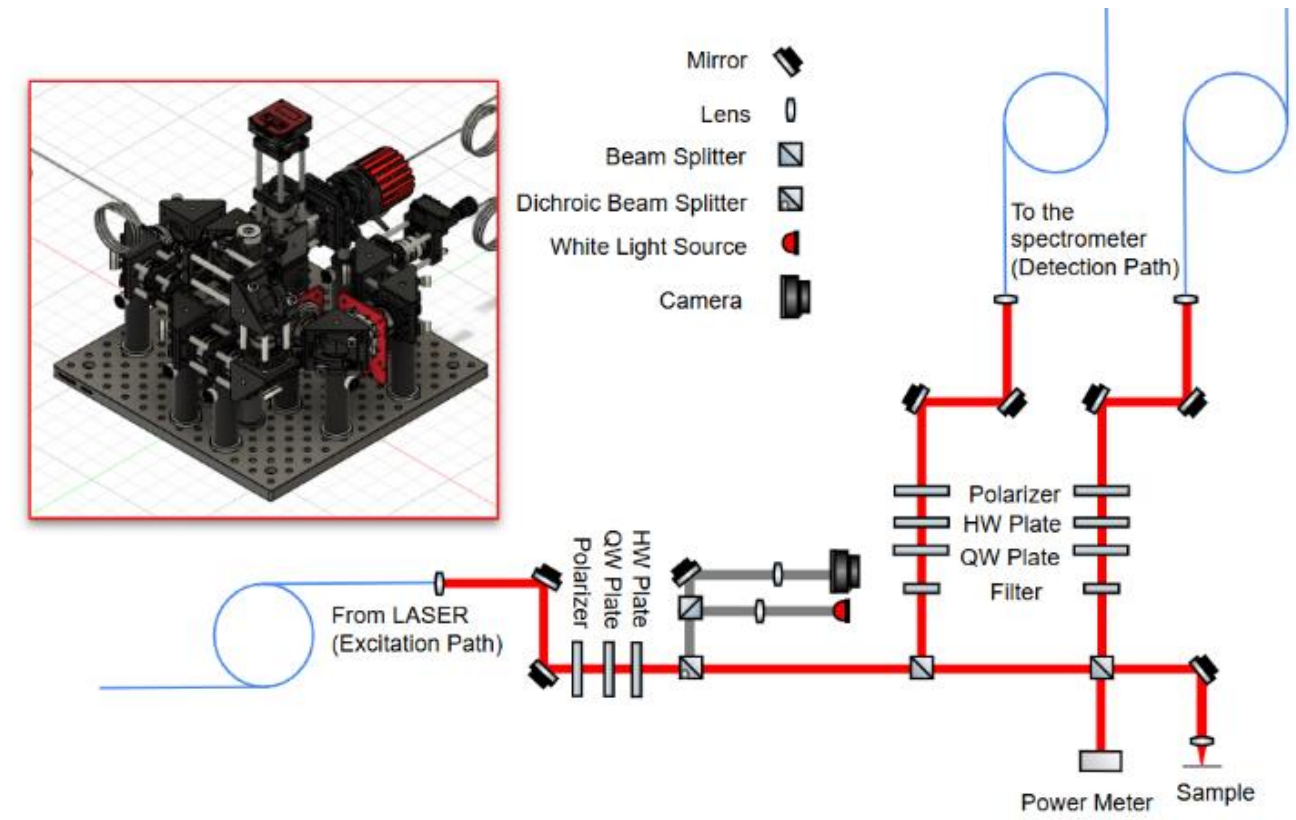

**Fig. 8.** A Schematic of an optical setup for efficient excitation and collection of the light from the cavity. inset: 3D configuration of the optical setup.

## References

[1] A. Einstein, B. Podolsky, and N. Rosen, "Can quantum-mechanical description of physical reality be considered complete?," *Physical Review*, vol. 47, no. 10, pp. 777–780, May 1935, doi: 10.1103/PHYSREV.47.777/FIGURE/1/THUMB.

[2] E. Schrödinger, "Discussion of Probability Relations between Separated Systems," *Mathematical Proceedings of the Cambridge Philosophical Society*, vol. 31, no. 4, pp. 555–563, 1935, doi: DOI: 10.1017/S0305004100013554.

[3] J. S. Bell, "On the Einstein Podolsky Rosen paradox," *Physics Physique Fizika*, vol. 1, no. 3, p. 195, Nov. 1964, doi: 10.1103/PhysicsPhysiqueFizika.1.195.

[4] A. Aspect, P. Grangier, and G. Roger, "Experimental Tests of Realistic Local Theories via Bell's Theorem," *Phys Rev Lett*, vol. 47, no. 7, p. 460, Aug. 1981, doi: 10.1103/PhysRevLett.47.460.

[5] S. J. Freedman and J. F. Clauser, "Experimental Test of Local Hidden-Variable Theories," *Phys Rev Lett*, vol. 28, no. 14, p. 938, Apr. 1972, doi: 10.1103/PhysRevLett.28.938.

[6] D. Bouwmeester, J. W. Pan, K. Mattle, M. Eibl, H. Weinfurter, and A. Zeilinger, "Experimental quantum teleportation," *Nature 1997 390:6660*, vol. 390, no. 6660, pp. 575–579, Dec. 1997, doi: 10.1038/37539.

[7] C. H. Bennett and G. Brassard, "Public key distribution and coin tossing," in *Proceedings of the IEEE International Conference on Computers, Systems, and Signal Processing, Bangalore (IEEE, New York, 1984)*, 1984, p. 175.

[8] P. W. Shor, "Algorithms for quantum computation: Discrete logarithms and factoring," *Proceedings - Annual IEEE Symposium on Foundations of Computer Science, FOCS*, pp. 124–134, 1994, doi: 10.1109/SFCS.1994.365700.

[9] R. Raussendorf and H. J. Briegel, "A One-Way Quantum Computer," *Phys Rev Lett*, vol. 86, no. 22, p. 5188, May 2001, doi: 10.1103/PhysRevLett.86.5188.

[10] J. Huang, X. Chen, X. Li, and J. Wang, "Chip-based photonic graph states," *AAPPS Bulletin*, vol. 33, no. 1, pp. 1–8, Dec. 2023, doi: 10.1007/S43673-023-00082-7/FIGURES/2.

[11] J. Huang, X. Chen, X. Li, and J. Wang, "Chip-based photonic graph states," *AAPPS Bulletin*, vol. 33, no. 1, pp. 1–8, Dec. 2023, doi: 10.1007/S43673-023-00082-7/FIGURES/2.

[12] H. J. Briegel, D. E. Browne, W. Dür, R. Raussendorf, and M. Van Den Nest, "Measurement-based quantum computation," *Nature Physics 2009 5:1*, vol. 5, no. 1, pp. 19–26, Jan. 2009, doi: 10.1038/nphys1157.

[13] D. Cogan, Z. E. Su, O. Kenneth, and D. Gershoni, "Deterministic generation of indistinguishable photons in a cluster state," *Nature Photonics 2023 17:4*, vol. 17, no. 4, pp. 324–329, Feb. 2023, doi: 10.1038/s41566-022-01152-2.

[14] I. Schwartz *et al.*, "Deterministic generation of a cluster state of entangled photons," *Science (1979)*, vol. 354, no. 6311, pp. 434–437, Oct. 2016, doi: 10.1126/SCIENCE.AAH4758/SUPPL_FILE/SCHWARTZ-SM.PDF.

[15] P. Thomas, L. Ruscio, O. Morin, and G. Rempe, "Efficient generation of entangled multiphoton graph states from a single atom," *Nature 2022 608:7924*, vol. 608, no. 7924, pp. 677–681, Aug. 2022, doi: 10.1038/s41586-022-04987-5.

[16] C. W. Chou, H. De Riedmatten, D. Felinto, S. V. Polyakov, S. J. Van Enk, and H. J. Kimble, "Measurement-induced entanglement for excitation stored in remote atomic ensembles," *Nature 2005 438:7069*, vol. 438, no. 7069, pp. 828–832, Dec. 2005, doi: 10.1038/nature04353.

[17] S. C. Benjamin, B. W. Lovett, and J. M. Smith, "Prospects for measurement-based quantum computing with solid state spins," *Laser Photon Rev*, vol. 3, no. 6, pp. 556–574, Nov. 2009, doi: 10.1002/LPOR.200810051.

[18] H. J. Briegel, D. E. Browne, W. Dür, R. Raussendorf, and M. Van Den Nest, "Measurement-based quantum computation," *Nature Physics 2009 5:1*, vol. 5, no. 1, pp. 19–26, Jan. 2009, doi: 10.1038/nphys1157.

[19] D. Istrati *et al.*, "Sequential generation of linear cluster states from a single photon emitter," *Nature Communications 2020 11:1*, vol. 11, no. 1, pp. 1–8, Oct. 2020, doi: 10.1038/s41467-020-19341-4.

[20] D. Cogan, Z. E. Su, O. Kenneth, and D. Gershoni, "Deterministic generation of indistinguishable photons in a cluster state," *Nature Photonics 2023 17:4*, vol. 17, no. 4, pp. 324–329, Feb. 2023, doi: 10.1038/s41566-022-01152-2.

[21] P. Hilaire, E. Barnes, and S. E. Economou, "Resource requirements for efficient quantum communication using all-photonic graph states generated from a few matter qubits," *Quantum*, vol. 5, p. 397, Feb. 2021, doi: 10.22331/q-2021-02-15-397.

[22] E. Knill, R. Laflamme, and G. J. Milburn, "A scheme for efficient quantum computation with linear optics," *Nature 2001 409:6816*, vol. 409, no. 6816, pp. 46–52, Jan. 2001, doi: 10.1038/35051009.

[23] N. H. Lindner and T. Rudolph, "Proposal for pulsed On-demand sources of photonic cluster state strings," *Phys Rev Lett*, vol. 103, no. 11, p. 113602, Sep. 2009, doi: 10.1103/PHYSREVLETT.103.113602/CSMG_SUPPLEMENTA_FINAL.PDF.

[24] Y. Shi and E. Waks, "Deterministic generation of multidimensional photonic cluster states using time-delay feedback," *Phys Rev A (Coll Park)*, vol. 104, no. 1, p. 013703, Jul. 2021, doi: 10.1103/PHYSREVA.104.013703/FIGURES/11/MEDIUM.

[25] N. H. Lindner and T. Rudolph, "Proposal for pulsed On-demand sources of photonic cluster state strings," *Phys Rev Lett*, vol. 103, no. 11, p. 113602, Sep. 2009, doi: 10.1103/PHYSREVLETT.103.113602/CSMG_SUPPLEMENTA_FINAL.PDF.

[26] B. D. Gerardot *et al.*, “Optical pumping of a single hole spin in a quantum dot,” *Nature 2008 451:7177*, vol. 451, no. 7177, pp. 441–444, Jan. 2008, doi: 10.1038/nature06472.

[27] I. Schwartz *et al.*, “Deterministic generation of a cluster state of entangled photons,” *Science (1979)*, vol. 354, no. 6311, pp. 434–437, Oct. 2016, doi: 10.1126/SCIENCE.AAH4758/SUPPL_FILE/SCHWARTZ-SM.PDF.

[28] A. Vezvaee, P. Hilaire, M. F. Doty, and S. E. Economou, “Deterministic Generation of Entangled Photonic Cluster States from Quantum Dot Molecules,” *Phys Rev Appl*, vol. 18, no. 6, p. L061003, Dec. 2022, doi: 10.1103/PHYSREVAPPLIED.18.L061003/FIGURES/3/MEDIUM.

[29] D. Cogan, Z.-E. Su, O. Kenneth, and D. Gershoni, “A deterministic source of indistinguishable photons in a cluster state,” Oct. 2021, Accessed: Feb. 13, 2025. [Online]. Available: https://arxiv.org/abs/2110.05908v1

[30] V. S. Ferreira, G. Kim, A. Butler, H. Pichler, and O. Painter, “Deterministic generation of multidimensional photonic cluster states with a single quantum emitter,” *Nature Physics 2024 20:5*, vol. 20, no. 5, pp. 865–870, Feb. 2024, doi: 10.1038/s41567-024-02408-0.

[31] N. Coste *et al.*, “High-rate entanglement between a semiconductor spin and indistinguishable photons,” *Nature Photonics 2023 17:7*, vol. 17, no. 7, pp. 582–587, Apr. 2023, doi: 10.1038/s41566-023-01186-0.

[32] H. Singh, D. Farfurnik, Z. Luo, A. S. Bracker, S. G. Carter, and E. Waks, “Optical Transparency Induced by a Largely Purcell Enhanced Quantum Dot in a Polarization-Degenerate Cavity,” *Nano Lett*, vol. 22, no. 19, pp. 7959–7964, Oct. 2022, doi: 10.1021/ACS.NANOLETT.2C03098/ASSET/IMAGES/LARGE/NL2C03098_0004.JPEG.

[33] M. Minkov, V. S.-S. reports, and undefined 2014, “Automated optimization of photonic crystal slab cavities,” *nature.comM Minkov, V SavonaScientific reports, 2014•nature.com*, 2014, doi: 10.1038/srep05124.

[34] T. Asano and S. Noda, “Optimization of photonic crystal nanocavities based on deep learning,” *Optics Express, Vol. 26, Issue 25, pp. 32704-32717*, vol. 26, no. 25, pp. 32704–32717, Dec. 2018, doi: 10.1364/OE.26.032704.

[35] H. Singh, D. Farfurnik, Z. Luo, A. S. Bracker, S. G. Carter, and E. Waks, “Optical Transparency Induced by a Largely Purcell Enhanced Quantum Dot in a Polarization-Degenerate Cavity,” *Nano Lett*, vol. 22, no. 19, pp. 7959–7964, Oct. 2022, doi: 10.1021/ACS.NANOLETT.2C03098.